\documentclass[a4paper]{spie}  
\usepackage[dvipdfm]{graphicx}

\title{DIOS: the Diffuse Intergalactic Oxygen Surveyor} 

\author{T.~Ohashi\supit{a}, M.~Ishida\supit{a}, S.~Sasaki\supit{a}, Y.~Ishisaki\supit{a}, 
K.~Mitsuda\supit{b}, N.~Y.~Yamasaki\supit{b}, R.~Fujimoto\supit{b}, Y.~Takei\supit{b}, 
Y.~Tawara\supit{c}, A.~Furuzawa\supit{c}, Y.~Suto\supit{d}, 
K.~Yoshikawa\supit{d}, H.~Kawahara\supit{d}, N.~Kawai\supit{e}, T.~G.~Tsuru\supit{f}, K.~Matsushita\supit{g}, T.~Kitayama\supit{h}
\skiplinehalf
\supit{a} Department of Physics, Tokyo Metropolitan University, 1-1 Minami-Ohsawa, Hachioji, Tokyo~192-0397, Japan\\
\supit{b} Institute of Space and Astronautical Science, Japan Aerospace Exploration Agency, 3-1-1, Yoshinodai, Sagamihara, Kanagawa~229-8510, Japan\\
\supit{c} Department of Astrophysics, Nagoya University, Furho-cho, Chikusa-ku, Nagoya~464-8602 , Japan\\
\supit{d} Department of Physics, University of Tokyo, 7-3-1 Hongo, Bunkyo-ku, Tokyo~113-0033, Japan\\
\supit{e} Department of Physics, Tokyo Institute of Technology, 2-12-1 Ookayama, Meguro-ku, Tokyo~152-8551, Japan\\
\supit{f} Department of Physics, Kyoto University, Kitashirakawa, Sakyo-ku, Kyoto~606-8502, Japan\\
\supit{g} Department of Physics, Tokyo University of Science, 1-3 Kagurazaka, Shinjuku-ku, Tokyo~162-8601, Japan\\
\supit{h} Department of Physics, Toho University, 2-2-1 Miyama, Funabashi, Chiba~274-8510, Japan\\
}

\authorinfo{Further author information: (Send correspondence to T. Ohashi.)\\T. Ohashi: E-mail: ohashi@phys.metro-u.ac.jp, Telephone: +81-426-77-2492}

\begin{document} 
  \maketitle 

\begin{abstract}
 We present our proposal for a small X-ray mission DIOS (Diffuse
Intergalactic Oxygen Surveyor), consisting of a 4-stage X-ray
telescope and an array of TES microcalorimeters, cooled with
mechanical coolers, with a total weight of about 400 kg.  The mission
will perform survey observations of warm-hot intergalactic medium
using OVII and OVIII emission lines, with the energy coverage up to
1.5 keV\@.  The wide field of view of about $50'$ diameter, superior
energy resolution close to 2 eV FWHM, and very low background will
together enable us a wide range of science for diffuse X-ray sources.
We briefly describe the design of the satellite, performance of the
subsystems and the expected results.
\end{abstract}



\section{INTRODUCTION}
\label{sect:intro}  

The microcalorimeter experiment XRS on-board Suzaku (Astro-E2), which
was launched in July 2005, has suffered from the loss of liquid helium
after about a month from the launch. However, the instrument was
cooled down to 60 mK and the energy resolution of 7 eV was obtained in
the orbit for about 2 weeks. This is a significant achievement and
proves that the XRS team has developed enough technologies to carry out
the microcalorimeter experiment in space. There is no doubt that
microcalorimeters will be the key instrument in the future X-ray
astronomy.

The study of warm-hot intergalactic medium (WHIM) is the remaining
frontier of X-ray astronomy. The importance of the observational study
of WHIM can be summarized as follows.
\begin{enumerate}
\item It is certain that baryons carry 4\% of the energy density in
the universe, and measurements of Lyman $\alpha$ forests give
consistent results. However, less than half of the baryonic matter is
directly probed in the present universe \cite{fukugita98}. Numerical
simulations predict that WHIM carries about 50\% of baryons. Direct
detection of WHIM will solve the problem of missing baryon, and the
cosmic baryon budget will be fully understood.

\item WHIM is the best tracer of the large-scale structure of the
universe. Galaxies, by optical surveys, distribute rather sparsely in
probing the continuous large-scale structure and clusters of galaxies,
by X-rays, only show us the densest part of the filamentary
structure. WHIM would reveal the faint part of the filament and enable
us to see the structure of dark matter directly.

\item WHIM is also an important probe for the thermal history of the
universe. WHIM has been ionized, heated and metal-enriched through the
past star/galaxy/structure-formation processes, therefore thermal and
chemical properties of the WHIM and their evolution with redshift
would give us unique information in looking into the thermal history
of the universe.

\end{enumerate} 

WHIM has been detected as absorption lines in the UV and X-ray spectra
of bright AGNs \cite{tripp04}. Recent Chandra detection of the
redshifted oxygen absorption lines in the spectra of Mrk421 is the
most remarkable example \cite{nicastro05}. However, absorption study
does not give us full information on the spatial distribution of
WHIM\@.  It is needless to say that emission from the WHIM gives us an
entirely new information, in particular about the spatial structure
based on survey observations. Another point in measuring both WHIM
emission and absorption in the same region is that one can solve the
density ($n$)and actual size ($L$) of the cloud, because emission
scales as $n^2L^3$ and absorption as $nL$. This may turn out to be a
powerful method to look into the structure of distant gas clouds.

Since the emissivity scales as $n^2$, the continuum emission from the
WHIM will be completely masked by the strong thermal emission from the
Milky-Way hot gas which is brighter by about 2 orders of
magnitude. The only possible way to detect WHIM emission is through
the oxygen lines, since redshifted WHIM lines can be separated from
the strong Galactic lines with energy resolution better than several
eV\@.  The existing X-ray missions are not sensitive enough, since
grating spectrometers are inappropriate to cover a large sky region,
for example.  The observation of oxygen emission lines from the WHIM
is only possible with non dispersive device like
microcalorimeters. This is the reason why we propose a dedicated small
mission to explore the structure of diffuse intergalactic medium,
called DIOS (Diffuse Intergalactic Oxygen Surveyor), which we hope to
be launched around 2010.

As described below, this mission naturally provides a very low
background. We can expect very rich unique science from this mission
besides the survey of WHIM\@.

\section{Spacecraft}

The view of the DIOS spacecraft is shown in Fig.\ \ref{spacecraft} and
main parameters of the satellite are listed in Table 1. The spacecraft
will weigh about 400 kg, out of which the payload takes $\sim 280$
kg. This mass enables the satellite to be launched as a piggy-back or
sub-payload in H2 or Ariane rockets. It is also possible for a launch
with the new ISAS rocket, such as M-V light. The size in the launch
configuration is $1.5 \times 1.5 \times 1.2$ m, and one side will be
expanded to about 6 m after the paddle deployment.

\begin{figure}[!htb] 
\begin{minipage}{0.48\textwidth}
\includegraphics[width=0.95\textwidth]{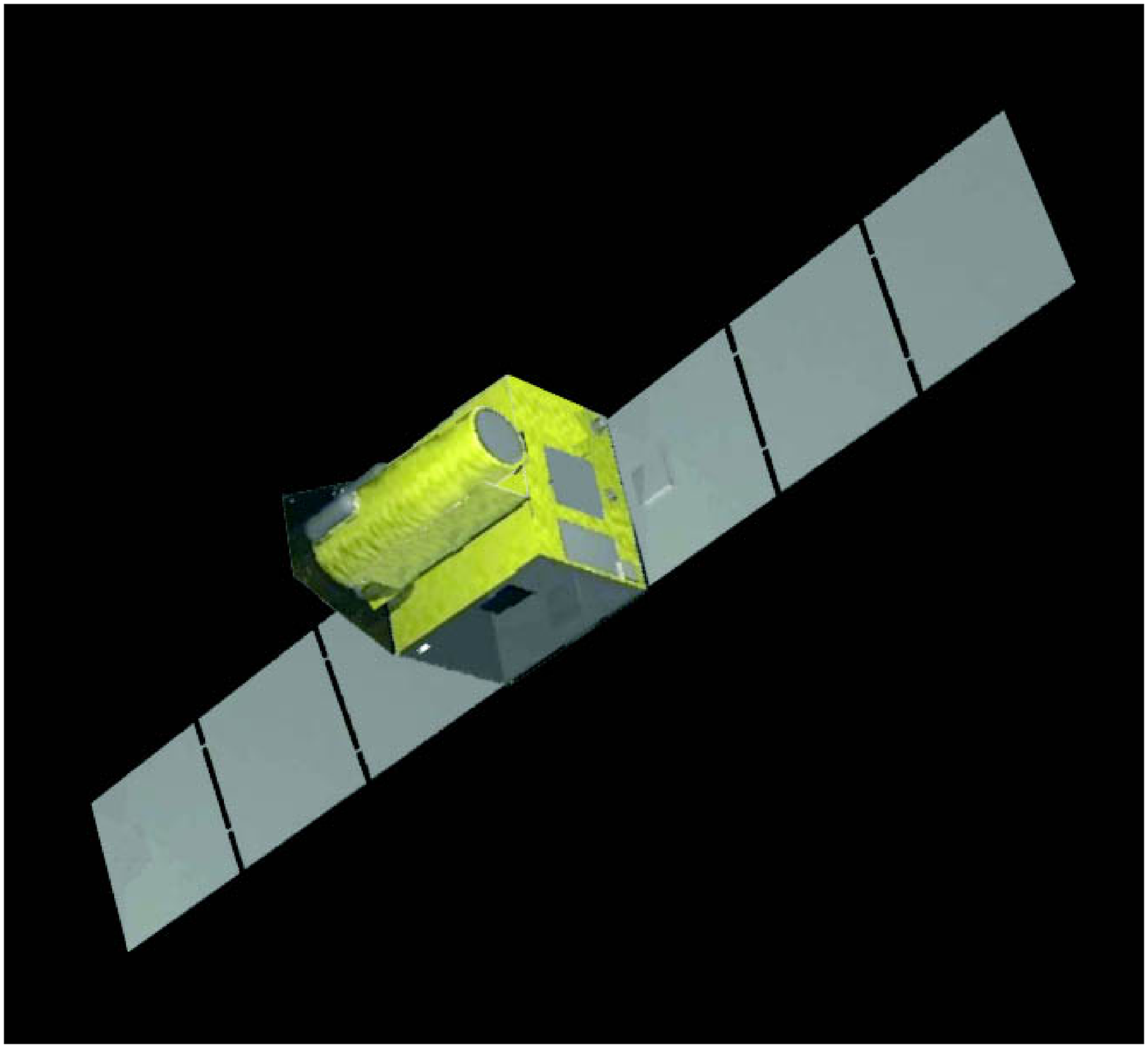}
\caption{The DIOS spacecraft. The length of the solar paddle is 6 m.}
\label{spacecraft}
\end{minipage}
\newcounter{keepfignum}
\setcounter{keepfignum}{\value{figure}}
\renewcommand{\figurename}{Table}
\setcounter{figure}{\value{table}}
\addtocounter{table}{1}
\begin{minipage}{0.5\textwidth}
\caption{Parameters of the observing instruments on board DIOS}
\begin{tabular}{|l|l|}\hline
Effective area & $> 100$ cm$^2$ \\
Field of view & $50'$ diameter \\
$S\Omega$ & $\sim 100$ cm$^2$ deg$^2$ \\
Angular resolution & $3'\ (16\times 16$ pixels) \\
Energy resolution & $< 5$ eV (FWHM) \\
Energy range & 0.1-- 1.5 keV \\
Observing life & $> 5$ yr \\ \hline
\end{tabular} 
\end{minipage}
\end{figure}
\renewcommand{\figurename}{Figure}
\setcounter{figure}{\value{keepfignum}}

The total power is 500 W, of which 300 W is consumed by the
payload. The nominal orbit is a near-earth circular one with an
altitude of 550 km, in the case of launch with the ISAS rocket. An
alternative choice of the orbit under consideration is an eccentric
geostationary transfer orbit. This orbit gives a lower heat input from
the earth and relaxes the thermal design of the satellite, and would
enhance the launch opportunity to be carried as a sub-payload with
geostationary satellites. One significant problem is the increased
particle background as recognized with Chandra and XMM-Newton. In the
soft energy range below 1 keV, electrons can be a major source of
background.

\begin{figure}[!tb] 
\begin{minipage}{0.49\textwidth}
\includegraphics[width=0.95\textwidth]{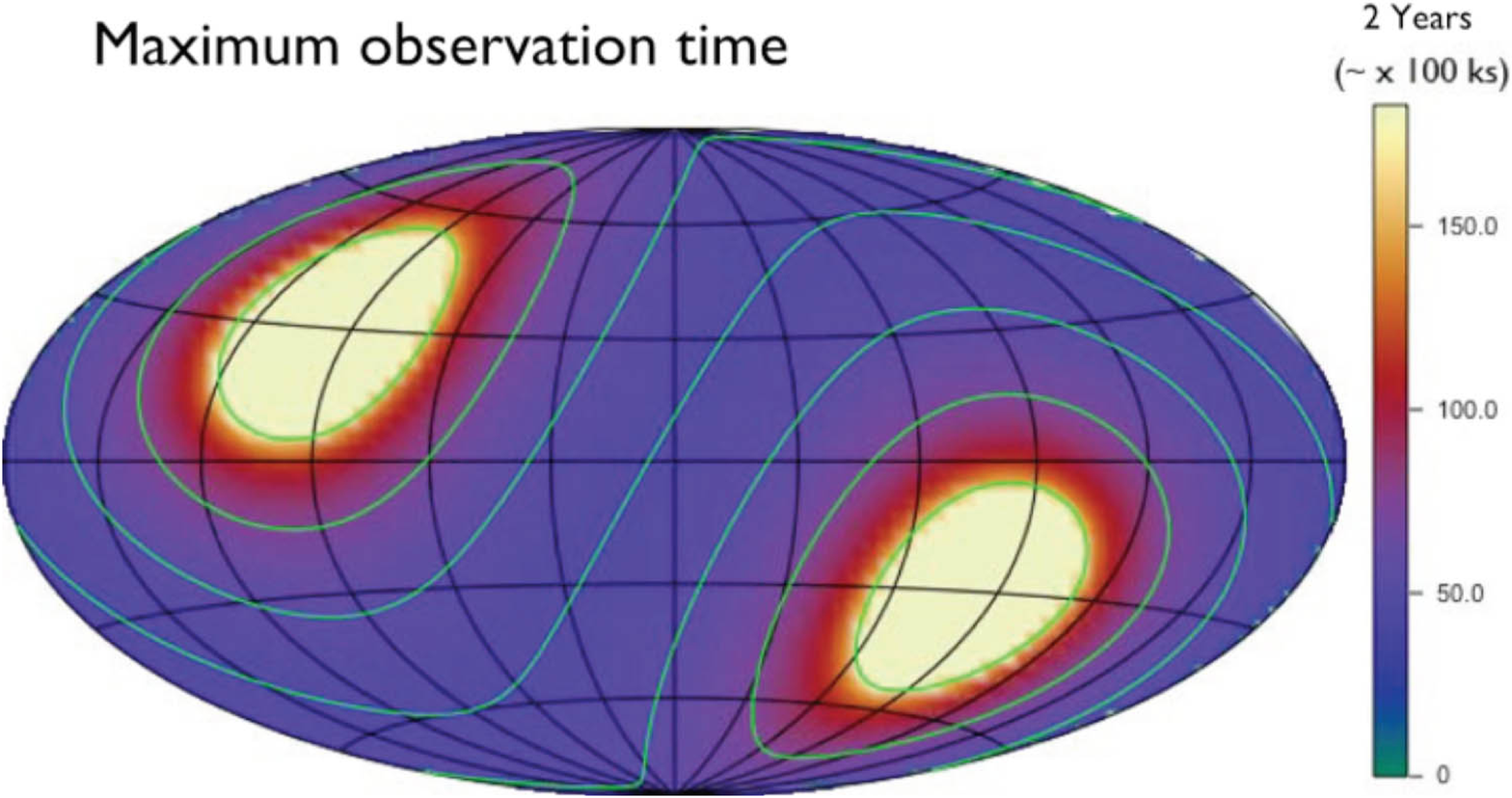}
\caption{Observable sky regions from DIOS. The ecliptic poles are
accessible throughout the year, and all the position can be observed
once in a half year period.} \label{observable_sky}
\end{minipage} \hfill
\begin{minipage}{0.48\textwidth}
\includegraphics[width=0.95\textwidth]{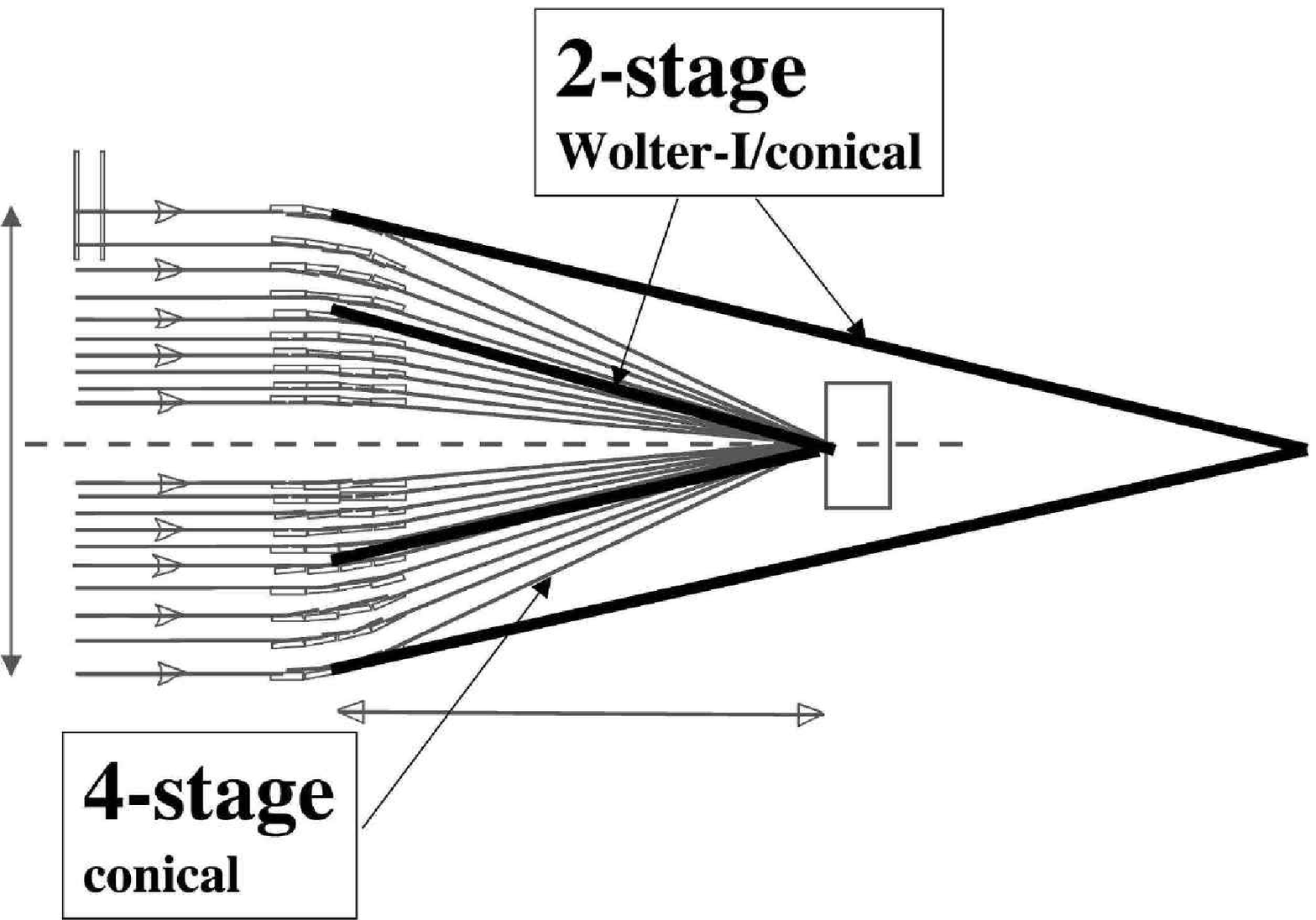}
\caption{Concept of the 4-stage reflection telescope.} \label{4reflection}
\end{minipage}
\end{figure}

The attitude will be 3-axis stabilized with momentum wheels. Typical
pointing accuracy will be about $10''$. The allowed range for the
direction of the field of view is $90^\circ \pm 25^\circ$ from the sun
direction. There is a radiator panel looking in the direction
perpendicular to both sun and the pointing directions. To avoid the
earth center covering the radiator direction, the pointing direction
of the satellite will be rotated by about $180^\circ$ in every
orbit. This attitude operation will take 6 minutes. Therefore, almost
opposite sky directions will be observed at the same time. With this
constraint, any position in the sky can be accessed within half a
year, and the ecliptic poles are accessible throughout the year. The
accessible sky regions and observable length in a year are shown in
Fig.\ \ref{observable_sky}. Clearly a large part of the sky can be
covered with DIOS, enabling not only the WHIM survey but also
large-scale mapping observations of the hot Galactic interstellar
medium.

\section{Instruments}

Several new technologies will be introduced in the DIOS mission. The
4-stage X-ray telescope FXT (Four-stage X-ray Telescope) is the first
new feature \cite{tawara04}. The details of the mirror development
activity will be reported in the same volume \cite{tawara06}. As shown
in Fig.\ \ref{4reflection}, incident X-rays are reflected 4 times by
thin-foil mirrors and are focused at about 70 cm from the mirror
level. The thin-foil technique was first used in BBXRT and then
successfully used for ASCA and Suzaku. For the oxygen lines at $\sim
0.6$ keV, the reflectivity of the mirror surface is close to 80\% and
the reduction of the effective area is not a serious problem. The
4-stage reflection gives the focal length about half as long as the
usual 2-stage design, and saves the volume and weight of the satellite
substantially. Also, a small focal plane detector can cover a
relatively wide field of view, which is a great advantage for the TES
calorimeter array.

\begin{figure}[!tb] 
\begin{minipage}{0.48\textwidth}
\includegraphics[width=0.95\textwidth,angle=0]{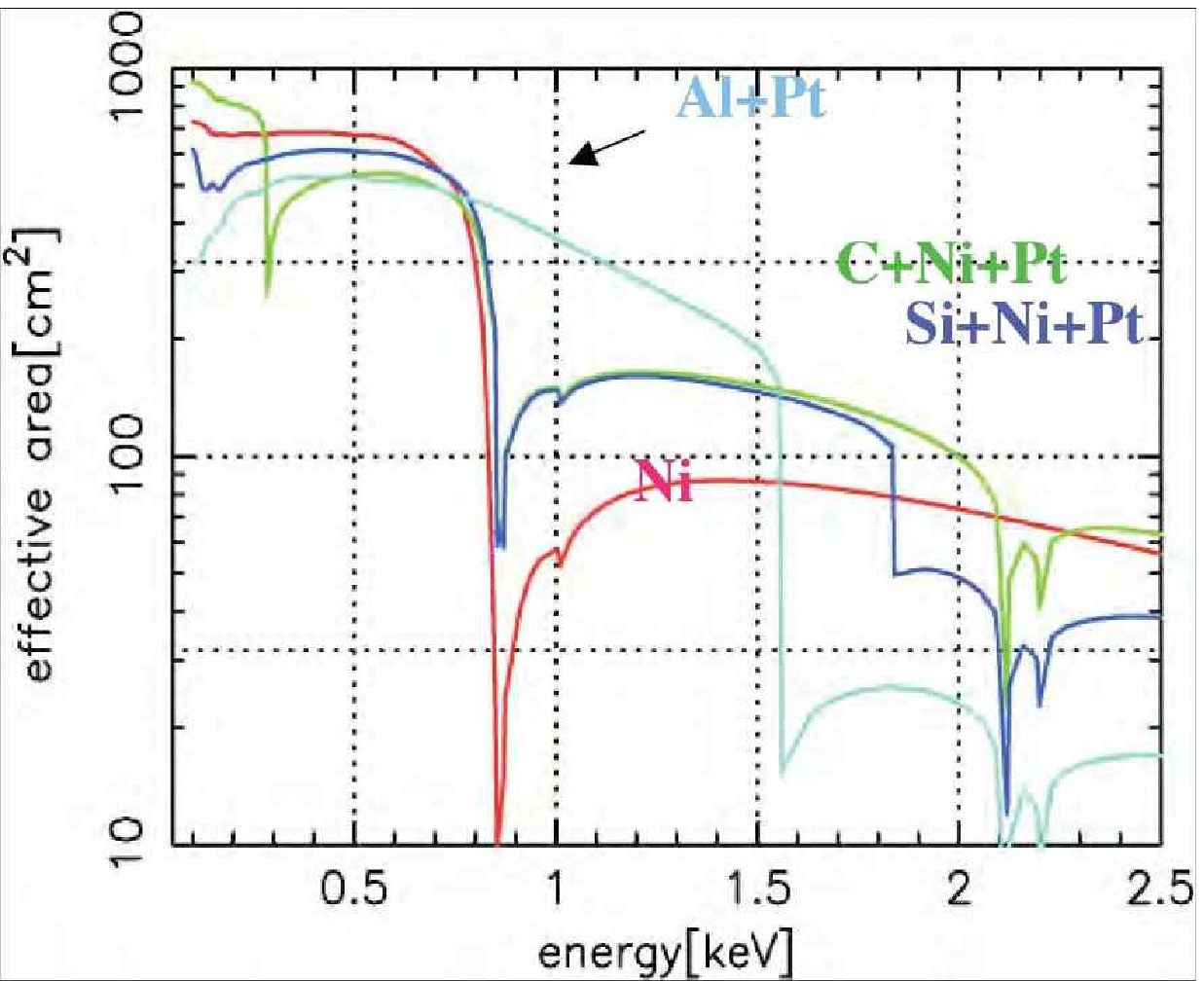}
\caption{Effective area as a function of energy for various multi-layer coatings. Al and Pt bilayer gives a good performance up to 1.5 keV\@. See Tawara et al.\ in this volume for details.} \label{AlPt}
\end{minipage} \hfill
\begin{minipage}{0.49\textwidth}
\includegraphics[width=0.98\textwidth]{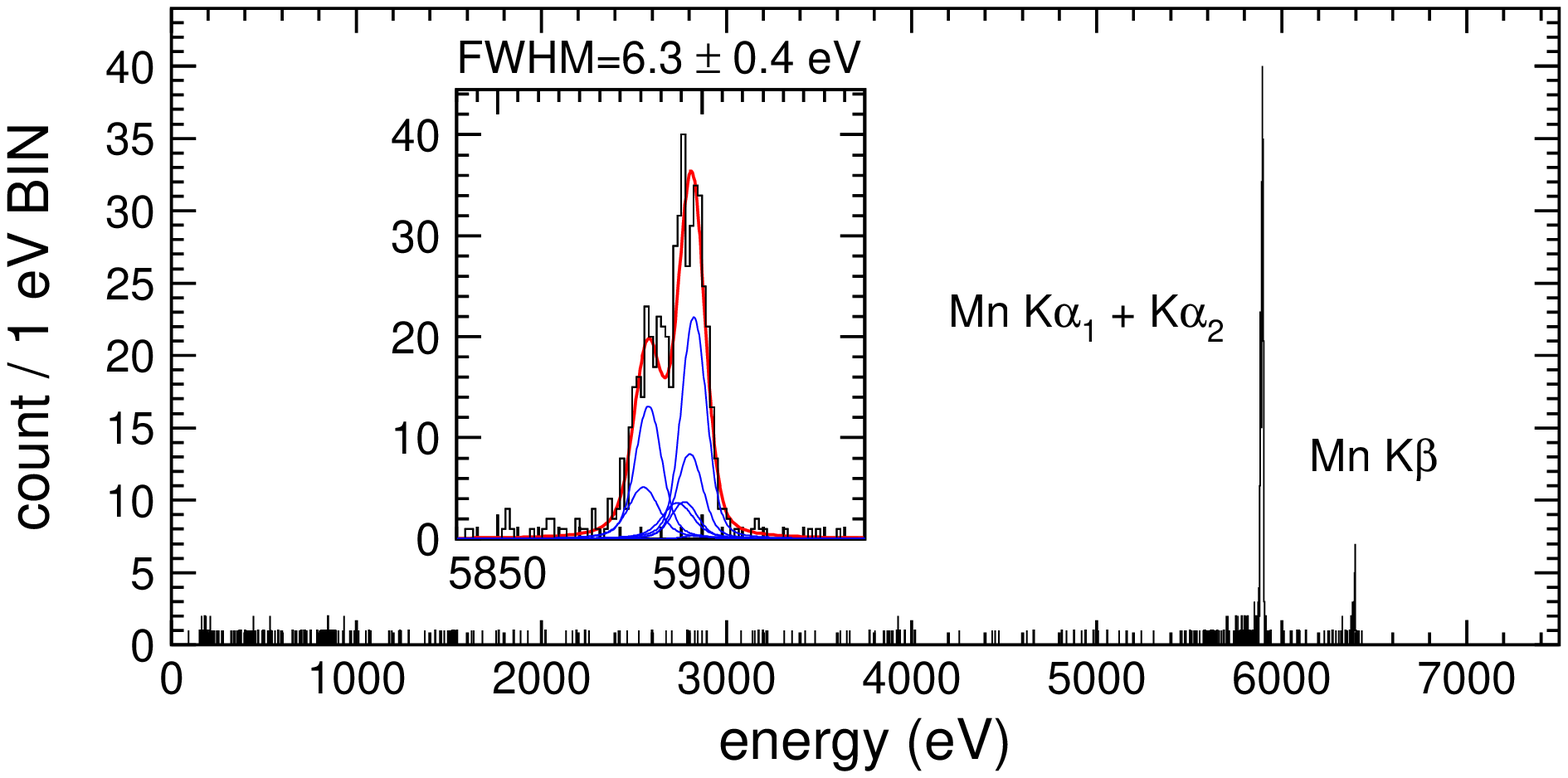}
\caption{Pulse-height spectrum for Mn-K$\alpha$ X-rays obtained by a single pixel Ti-Au TES microcalorimeter.} \label{phdist}
\end{minipage}
\end{figure}

In our basic design, the outer diameter of the mirror and the
effective area at 0.6 keV are 50 cm and 400 cm$^2$, respectively.
Ray-tracing simulation indicates that the angular resolution is $2'$
(half-power diameter), and the image quality does not show significant
degradation at an offset angle $30'$. Therefore, FXT really gives the
field of view of about 1 degree. The reflectivity will be optimized by
employing a multi-layer coating on the mirror surface. Al and Pt
bilayer will give a good reflectivity up to 1.5 keV (see Fig.\
\ref{AlPt}), which will enable us to observe Ne, Fe-L and Mg
lines. However, direct coating of Al and Pt does not work since this
combination does not make a stable multilayers. Appropriate choice of
the multilayer material will give an energy coverage up to 1.5 keV
(see Tawara et al.\cite{tawara06}\ for details). With this technique,
we can perform a spectral study of hot plasmas with a temperature up
to a few keV from the DIOS\@. Most of clusters of galaxies and
supernova remnants will be good targets.

Another advantage of the FXT design is the very low background. The
smaller detector size for a given angular size, as a result from the
short focal length, means that the internal background is very low. In
fact, the focal length is about 10 times shorter than the XMM-Newton
mirror, for example, indicating that 100 times smaller detector sees
the same sky region with roughly 10 times smaller effective area. The
low-earth orbit of DIOS also helps observation under the low
background condition as well. Therefore, if we include all these
factors, the background of DIOS would be more than 10 times lower than
the level of Chandra and XMM-Newton in observing the same sky
region. This makes DIOS a very powerful satellite to look at the WHIM
and other faint emission such as outer regions of clusters of
galaxies.

The focal plane instrument XSA (X-ray Spectrometer Array) is an array
of TES microcalorimeters, whose development in Japan is a
collaboration with Waseda University, Seiko Instruments Inc., and
Mitsubishi Heavy Industries. The array consists of $16\times 16$
pixels covering an area of about 1 cm square. The corresponding field
of view is $50'$. XSA will have an energy resolution better than 5 eV
FWHM at 0.6 keV\@. An example of the pulse-height spectra obtained for
5.9 keV X-rays with our TES calorimeter, which is a single pixel type
made with Ti-Au bilayer with a Au X-ray absorber, is shown in Fig.\
\ref{phdist}\cite{ishisaki03}. The measured resolution is 6.3 eV
FWHM\@. Since the energy range of XSA is below 1.5 keV, the required
heat capacity of the X-ray absorber is significantly low. Therefore, a
significantly better energy resolution will be achievable in the soft
X-ray range.

\begin{figure}[!] 
\hfill
\begin{minipage}{0.48\textwidth}
\includegraphics[width=0.95\textwidth]{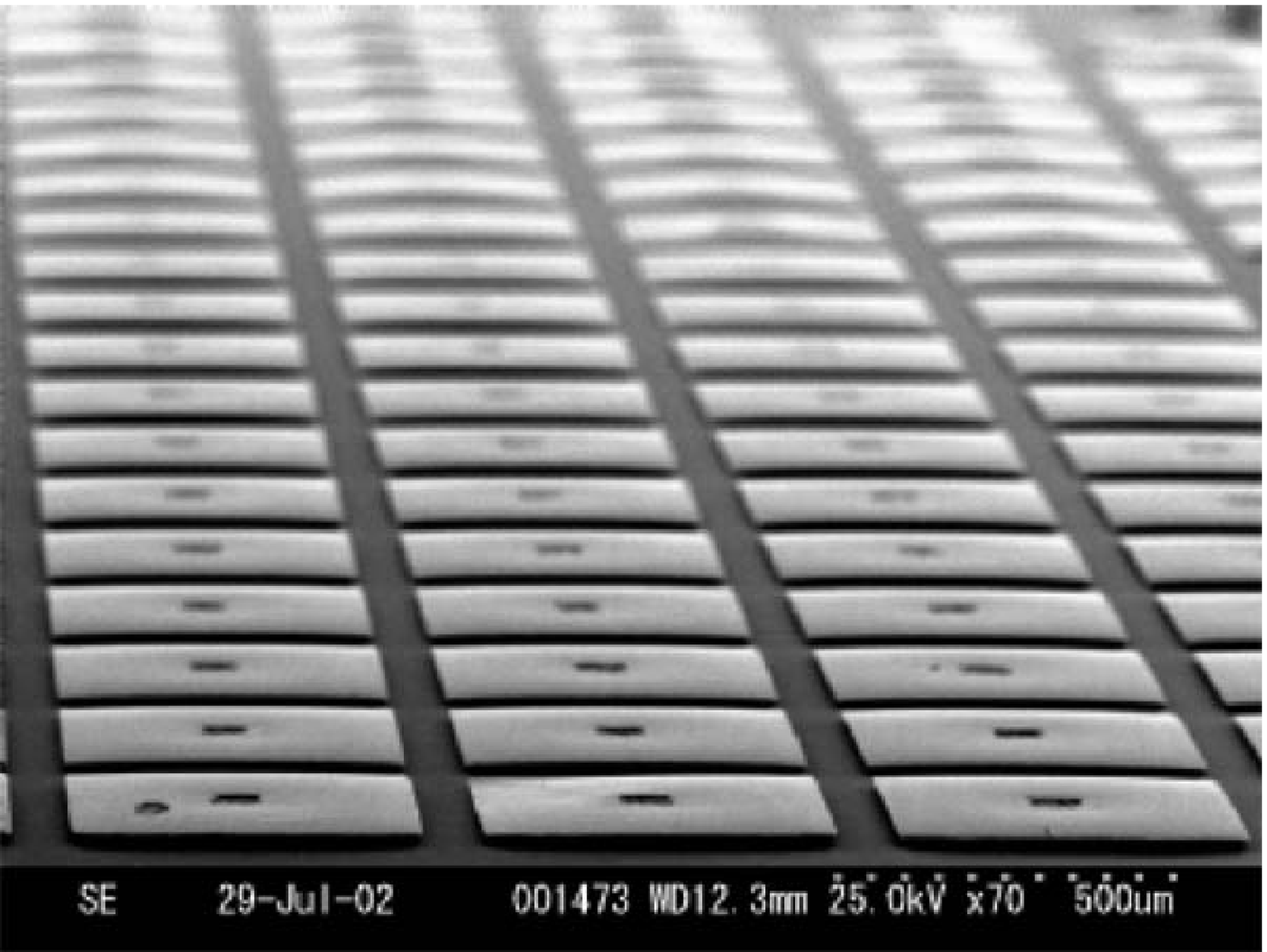}
\caption{Mechanical model of the 256 pixel absorber for the TES array, produced by a Sn plating. Size of one pixel is 500 $\mu$m.} \label{tesarray}
\end{minipage} \hfill
\begin{minipage}{0.48\textwidth}
\includegraphics[width=0.85\textwidth]{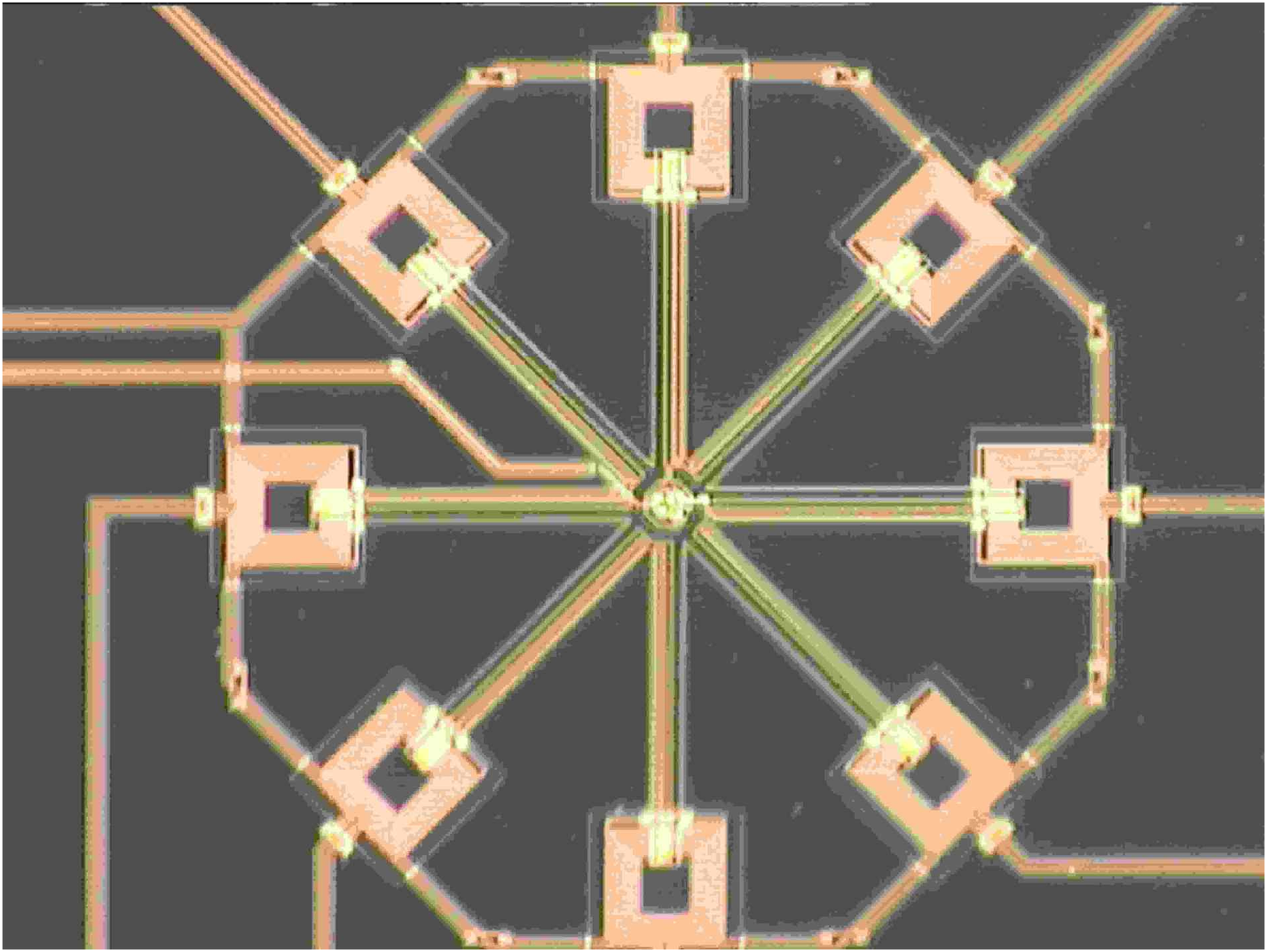}
\caption{Test model of the 8-input SQUID.} \label{8-input}
\end{minipage}
\end{figure}

We are now developing several new techniques toward the multi-pixel
operation of TES calorimeters \cite{ishisaki04,morita06}. X-ray
absorbing material under consideration is Bi, which has low heat
capacity and does not produce long-life quasi particles in the
absorber. We are testing an electro-plating method to build a $16
\times 16$ absorber array with a pixel size $\sim 0.5 \times 0.5$ mm
supported by a thin stem (see Fig.\ \ref{tesarray} for our test model
made with Sn). In this figure, all the pixels have wirings equipped
and they can all operate as microcalorimeters. Several pixels indeed
show X-ray sensitivity, however absorbers are not mechanically stable
under the thermal cycle. We are looking into other stress-free method
to mount the absorber structure.

For the signal readout, efficient multiplexing of the signals is
essential to take all the data out from the cold stage. We are trying
to add the signals, e.g.\ 8 of them, in frequency space by driving the
TES calorimeters with AC bias at different frequencies for different
pixels\cite{iyomoto04,yamasaki06}. The summed signals are retrieved by single
wire and then demodulated by a room-temperature electronics. The SQUID
needs to be operated at 1 MHz to obtain a dynamic range in the
frequency space. We have produced 8-input SQUID (see Fig.\
\ref{8-input}), in collaboration with the Seiko Instruments Inc., to
add the TES signals, and it is under a performance test now.  An
efficient thermal shield with high soft X-ray transmission is also an
essential part, and a layer of 5 parylene films with $0.1 \mu$m
thickness each has been successfully used in the laboratory.

\begin{figure}[!htb]\begin{center}
 \includegraphics[width=0.55\textwidth,angle=0]{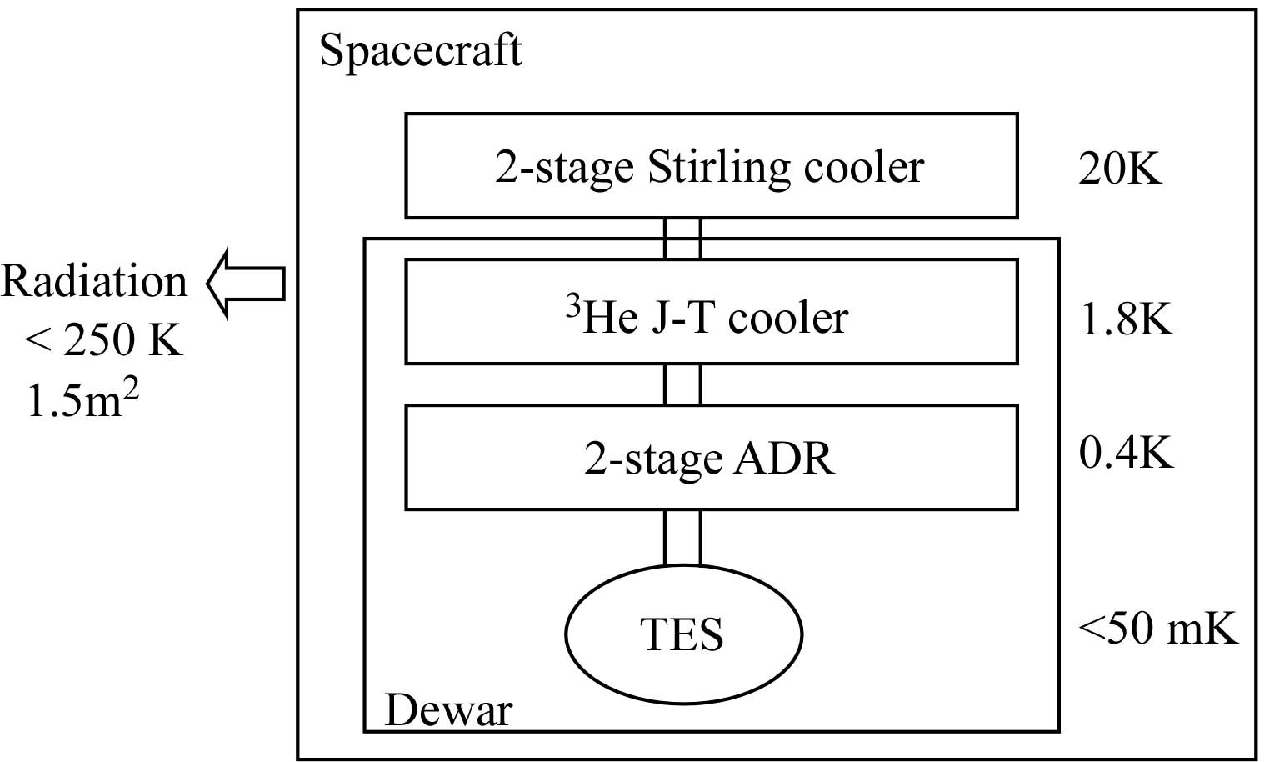}
 \caption{Cooling system for the TES calorimeter system.} \label{coolers}
\end{center} \end{figure}

Another important feature of DIOS is the cryogen-free cooling
system. As shown in Fig.\ \ref{coolers}, we are considering a serial
connection of different types of coolers to achieve the $\sim 50$ mK
operation for XSA within the available power budget. The outside panel
is equipped with a radiator which has $1.5\ {\rm m}^2$ area. The first
stage of the cooler is a 2-stage Stirling cooler which takes the
temperature down to 20 K, and then $^3$He Joule-Thomson cooler reduces
it to 1.8 K as the second stage. Then a 2-stage adiabatic
demagnetization refrigerator obtains the operating temperature at 50
mK\@. Since no cryogen is involved, this cooling system ensures an
unlimited observing life in the orbit, which is a significant
advantage of DIOS\@. The XSA system is subject to a warm launch,
therefore we have to allow for the initial cooling of the system for
the first 1--2 months in the orbit. We note that Stirling coolers are
operating successfully in Suzaku continuously for more than 10 months
in space. The coolers are under development with Sumitomo Heavy
Industry.

\section{Observation}

The main purpose of DIOS is a sky survey of WHIM using oxygen K
emission lines, based on the superior energy resolution and a large
field of view.  Yoshikawa et al.\cite{yoshikawa03,yoshikawa04}\
carried out detailed simulation for the spectra expected from the DIOS
observations. It is based on a cosmological hydrodynamic simulation
for a cube of $75h^{-1}$ Mpc size, which corresponds to an angular
size of $5^\circ$ at a redshift 0.3. The cosmological evolution of
intergalactic medium is included. The metallicity of the gas is
explicitly given, and the effective oxygen abundance is 0.1--0.2
solar. The WHIM emission in a redshift range $z=0-0.3$ is accumulated
along various trial lines of sight in the sky and the resultant X-ray
spectra are calculated. In this simulation, WHIM is assumed to be
under a collisional ionization equilibrium.

\begin{figure}[!htb]\begin{minipage}{0.68\textwidth}
\includegraphics[width=\textwidth]{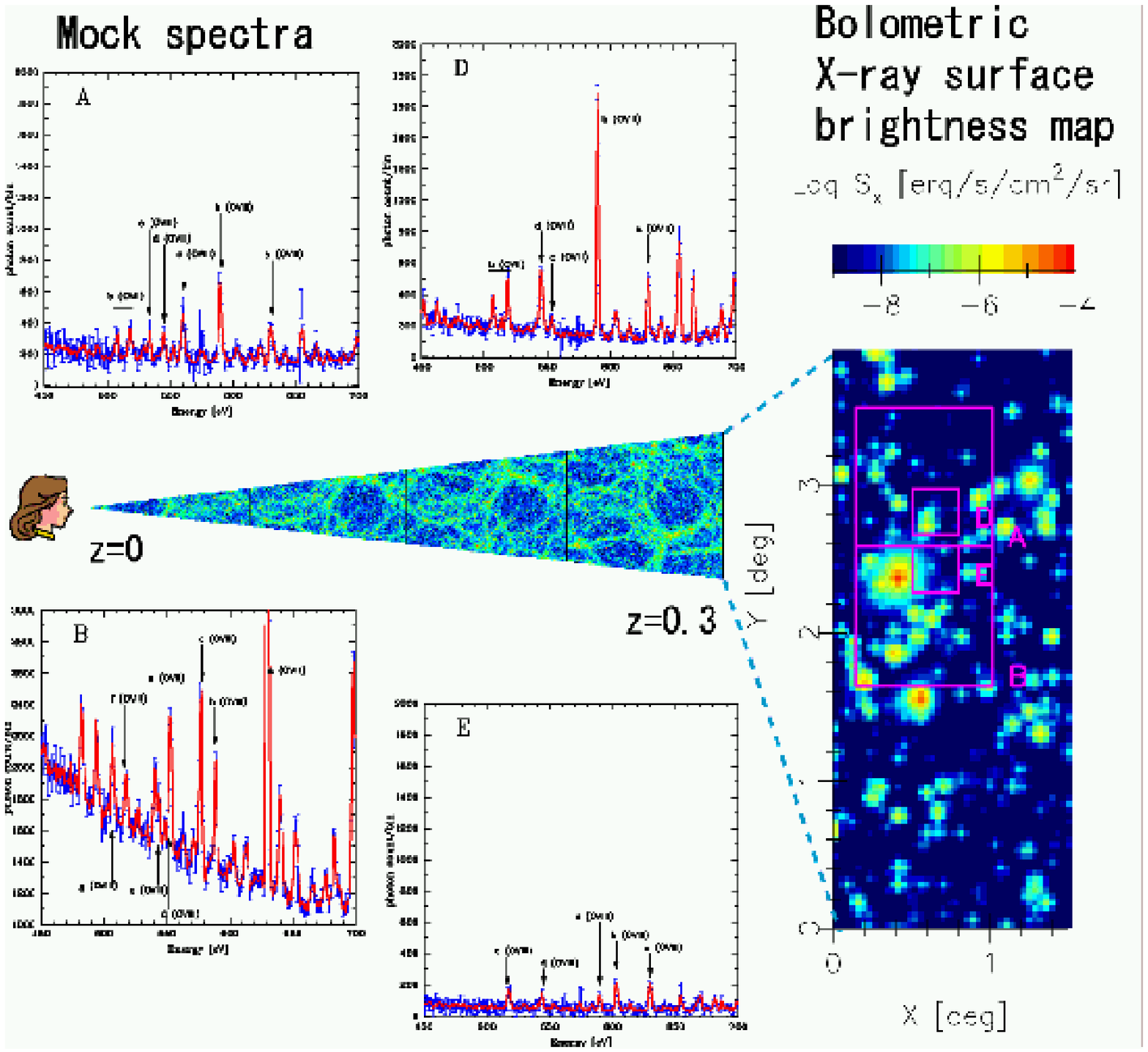}
\end{minipage} \hfill
\begin{minipage}{0.3\textwidth}
\caption{Simulated observations of WHIM observed from DIOS
\cite{yoshikawa03}. Typical observation time is 100 ksec. Foreground
Galactic emission is included in the simulation and subtracted from
the data.}\label{dios_outline}
\end{minipage}
\end{figure}

The simulation shows that each pointing will need typically 100 ksec
to obtain enough oxygen photons from the IGM\@. Fig.\
\ref{dios_outline} shows the expected pulse-height spectra for
different sky regions. The strong emission lines from our Galaxy are
included in the simulated observation and then subtracted in this
plot. So, the statistical fluctuation of the data includes this
effect. Strong oxygen lines are expected from relatively dense WHIM
regions, which are either close to clusters of galaxies or containing
groups of galaxies (such as the region B)\@.  Note that the hot
interstellar medium in our galaxy produces typically 100 times
stronger line emission, which can be separated from the redshifted
WHIM lines with the energy resolution of the XSA\@. When the data
quality is high, the lines in the OVII triplet are clearly resolved
and we will be able to measure the temperature directly. These 3 lines
can also be used to separate individual plasma components when several
emission regions in the same line of sight but different redshifts are
observed together.

\begin{figure}[!htb]\begin{center}
\includegraphics[width=0.8\textwidth]{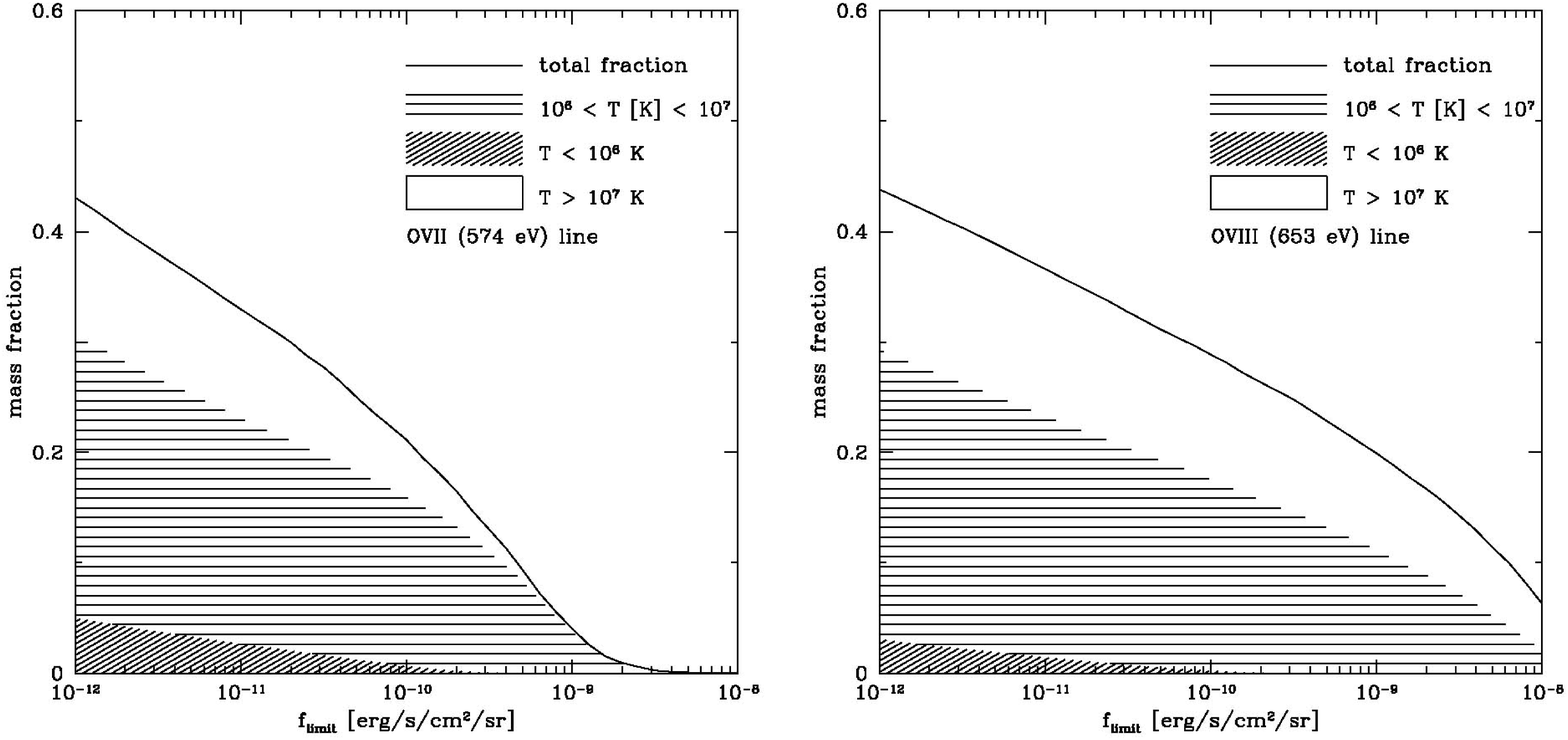}
\caption{Probed fraction of baryons plotted as a function of detection
limit \cite{yoshikawa04}. The left panel is for OVII line and the
right one for OVIII line, respectively.}
\label{probed_fraction}
\end{center} \end{figure}

The fraction of baryons probed by the detection limit of DIOS is shown
in Fig.\ \ref{probed_fraction}, in which mass fraction of baryons are
plotted as a function of detection flux. For a nominal detection limit
of DIOS, i.e.\ $10^{-10}-10^{-11}$ erg cm$^{-2}$ s$^{-1}$ sr$^{-1}$,
about 20\% of all the baryons will be detected. We note that OVIII
line is more efficient than OVII line, since the temperature range in
which the line is emitted is wider. Aside from the hot cluster gas
($T> 10^7$ K), DIOS can detect roughly half of the missing baryons,
namely the gas in the temperature range $10^6-10^7$ K\@. This
indicates that, even the DIOS satellite itself is quite small, the
mission can carry out the expected science.


The DIOS mission can also perform a mapping observation of the hot
interstellar medium in our galaxy. The high energy resolution ($\Delta
E < 5$ eV) will reveal the Doppler shifts of the hot interstellar gas
with a velocity $\sim 100$ km s$^{-1}$. The observations from DIOS can
confirm the large-scale outflow and falling-in of the hot bubbles in
our galaxy (galactic fountain).  Apart from this science, the energy
coverage up to 1.5 keV with superior resolution, combined with the
wide field of view ($50'$ diameter) and the extremely low internal
background, make DIOS a very powerful satellite for the study of
diffuse hot gas in various scales. Hot-gas properties near the virial
radius of clusters of galaxies remain highly uncertain even with the
Chandra and XMM-Newton observations. DIOS will give us a clear answer
about the temperature and metallicity in this region and will give new
constraints on the shock heating of the infalling gas. Also, study of
dynamical processes (shocks and bullets) in supernova remnants will
advance significantly with the high resolution spectroscopy from
DIOS\@. In this view, the capability of DIOS is highly complementary
to the existing and planned large X-ray observatories.

\begin{figure}[!htb] \begin{center}
\includegraphics[width=0.5\textwidth]{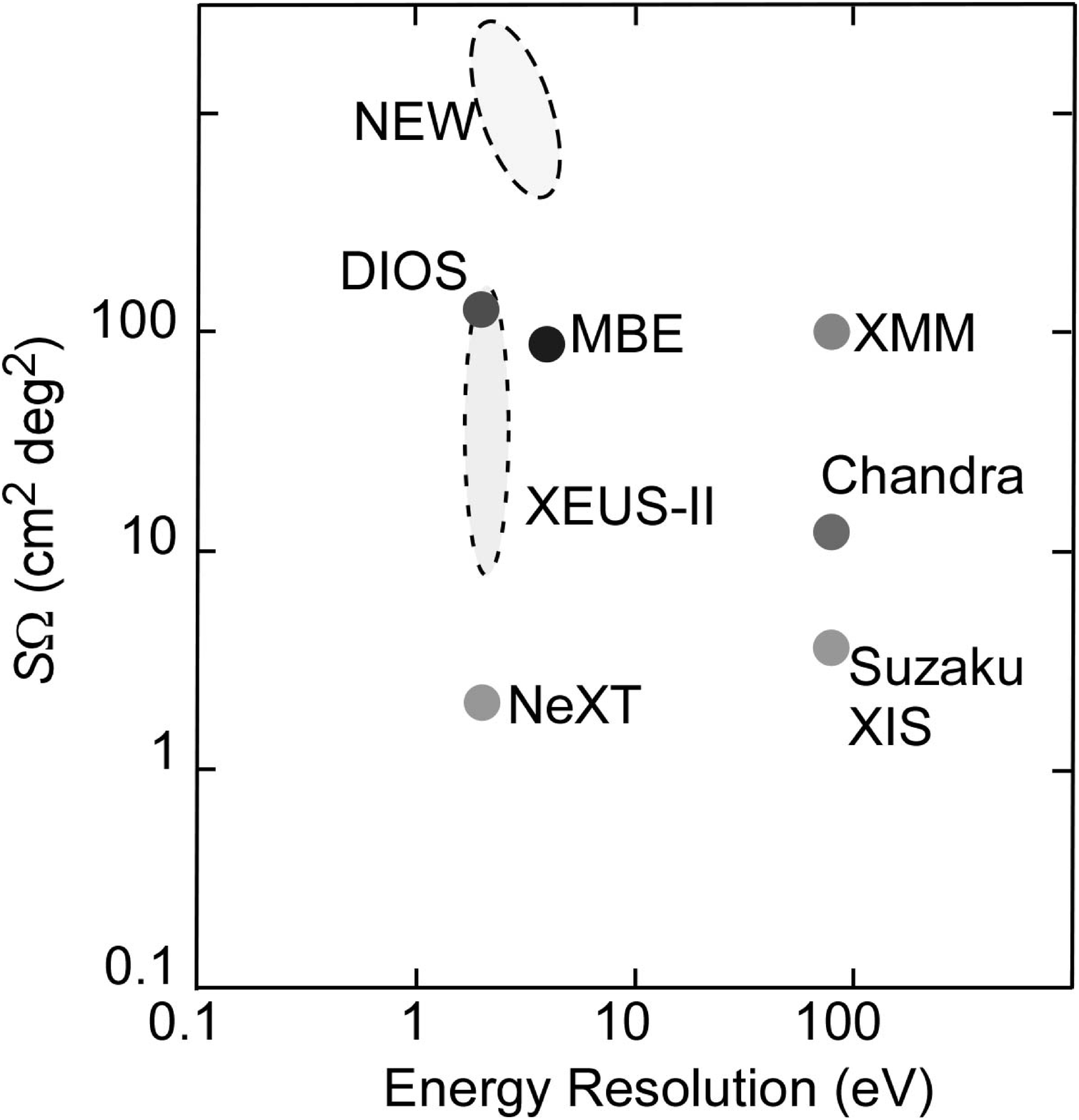}
\caption{Comparison of $S\Omega$ and energy resolution for
spectroscopic instruments (CCDs and microcalorimeters) for planned and
operating X-ray satellites.} \label{hikaku}
\end{center}
\end{figure}

Effective observing time will be approximately 40 ksec per day for a
near earth orbit. So if we spend 100 ksec in each pointing position, a
sky map for an area of $10^\circ \times 10^\circ$ can be produced in
about a year. This angular size is enough to see the large-scale
structure of the universe at $z < 0.3$. This survey observation of a
limited sky region will be the first task of DIOS\@. Since oxygen
lines from the galactic ISM are stronger by roughly 2 orders of
magnitude, 1 ksec in each point is enough to make a large map of the
ISM distribution. We plan to devote the second year for a survey of a
$100^\circ\times 100^\circ$ sky for the galactic ISM\@. After these 2
years, further deeper observations of the IGM as well as of various
selected objects can be performed. Fig.\ \ref{hikaku} compares
$S\Omega$, the grasp as the product of effective area and solid angle,
and energy resolution among different instruments in the X-ray
missions which are planned or already operating. Clearly, DIOS will
achieve the highest sensitivity for soft X-ray lines from extended
objects, except for the larger mission, NEW, based on the similar
concept.

\section{Prospect}

DIOS has been proposed to ISAS/JAXA in 2005 together with several tens
of other small missions. This year, 5 proposals were chosen as a pilot
example of small missions to be reported to JAXA headquarter. DIOS is
one of them, and only one for the astronomy mission.  Even though the
program of small missions has not started in JAXA, DIOS may have a
chance of fairly early opportunity. On the other hand, Japanese X-ray
group is proposing NeXT, which is a larger mission also carrying
microcalorimeters. In the case that the two mission might cause some
kind conflict, we would think of down sizing of the DIOS mission. At
the same time, we are exploring international collaboration. Dutch
mission NEW and Italian ESTREMO share the same scientific purpose with
DIOS\@. It is quite likely that hardware team would collaborate and
concentrate on the most promising single mission.

\bibliography{ohashi_spie06}   

\begin{thebibliography}{10}

\bibitem{fukugita98}
M.~Fukugita, C.~J. Hogan, and P.~J.~E. Peebles, ``The cosmic baryon budget,''
  {\em ApJ}~{\bf 503}, p.~518, 1998.

\bibitem{tripp04}
T.~M. Tripp, D.~V. Bowen, K.~R. Sembach, E.~B. Jenkins, B.~D. Savage, and
  P.~Richter, ``Missing baryons in the warm-hot intergalactic medium,'' ~{\bf
  astro-ph/0411151}, 2004.

\bibitem{nicastro05}
F.~{Nicastro}, S.~{Mathur}, M.~{Elvis}, J.~{Drake}, F.~{Fiore}, T.~{Fang},
  A.~{Fruscione}, Y.~{Krongold}, H.~{Marshall}, and R.~{Williams}, ``Chandra
  detection of the first x-ray forest along the line of sight to \rm{Markarian}
  421,'' {\em ApJ}~{\bf 629}, p.~700, 2005.

\bibitem{tawara04}
Y.~Tawara, Y.~Ogasaka, K.~Tamura, and A.~Furuzawa, ``Development of four-stage
  x-ray telescope for wide-field fine spectroscopic mission,'' {\em SPIE}~{\bf
  5168}, p.~386, 2004.

\bibitem{tawara06}
Y.~Tawara, A.~Furuzawa, Y.~Ogasaka, R.~Shibata, and K.~Tamura, ``The
  demonstration model of four-stage \rm{X}-ray telescope for \rm{DIOS},'' {\em
  SPIE}~{\bf 6266}, 2006.

\bibitem{ishisaki03}
Y.~{Ishisaki}, U.~{Morita}, T.~{Koga}, K.~{Sato}, T.~{Ohashi}, K.~{Mitsuda},
  N.~Y. {Yamasaki}, R.~{Fujimoto}, N.~{Iyomoto}, T.~{Oshima}, K.~{Futamoto},
  Y.~{Takei}, T.~{Ichitsubo}, T.~{Fujimori}, S.~{Shoji}, H.~{Kudo},
  T.~{Nakamura}, T.~{Arakawa}, T.~{Osaka}, T.~{Homma}, H.~{Sato},
  H.~{Kobayashi}, K.~{Mori}, K.~{Tanaka}, T.~{Morooka}, S.~{Nakayama},
  K.~{Chinone}, Y.~{Kuroda}, M.~{Onishi}, and K.~{Otake}, ``Present performance
  of a single pixel \rm{Ti/Au} bilayer \rm{TES} calorimeter,'' {\em SPIE}~{\bf
  4851}, p.~831, 2003.

\bibitem{ishisaki04}
Y.~{Ishisaki}, T.~{Ohashi}, T.~{Oshima}, U.~{Morita}, K.~{Shinozaki},
  K.~{Sato}, K.~{Mitsuda}, N.~Y. {Yamasaki}, R.~{Fujimoto}, Y.~{Takei},
  H.~{Sato}, N.~{Takahashi}, T.~{Homma}, and T.~{Osaka}, ``Development of a
  microcalorimeter array for the diffuse intergalactic oxygen surveyor
  (\rm{DIOS}) mission,'' {\em SPIE}~{\bf 5501}, p.~123, 2004.

\bibitem{morita06}
U.~Morita, Y.~{Yamakawa}, T.~{Fujimori}, Y.~{Ishisaki}, T.~{Ohashi},
  Y.~{Takei}, K.~{Yoshida}, T.~{Yoshino}, K.~{Mitsuda}, N.~Y. {Yamasaki},
  R.~{Fujimoto}, H.~{Sato}, Y.~{Minoura}, N.~{Takahashi}, T.~{Homma},
  S.~{Shoji}, Y.~{Kuroda}, and M.~{Onishi}, ``Evaluation of 256-pixel tes
  microcalorimeter arrays with electrodeposited \rm{Bi} absorbers,'' {\em
  Nucl.\ Instr.\ and Meth.}~{\bf A 559}, p.~539, 2006.

\bibitem{iyomoto04}
N.~Iyomoto, T.~Ichitsubo, K.~Mitsuda, N.~Y. Yamasaki, R.~Fujimoto, T.~Oshima,
  K.~Futamoto, Y.~Takei, T.~Fujimori, K.~Yoshida, Y.~Ishisaki, U.~Morita,
  T.~Koga, K.~Shinozaki, K.~Sato, N.~Takai, T.~Ohashi, T.~Miyazaki,
  S.~Nakayama, K.~Tanaka, T.~Morooka, and K.~Chinone, ``Frequency-domain
  multiplexing of tes microcalorimeter array with \rm{CABBAGE},'' {\em Nucl.\
  Instr.\ and Meth.}~{\bf A 520}, p.~566, 2004.

\bibitem{yamasaki06}
N.~Y. {Yamasaki}, K.~Masui, K.~Mitsuda, T.~Morooka, S.~Nakayama, and Y.~Takei,
  ``Design of frequency domain multiplexing of \rm{TES} signals by multi-input
  \rm{SQUIDs},'' {\em Nucl.\ Instr.\ and Meth.}~{\bf A 559}, p.~790, 2006.

\bibitem{yoshikawa03}
K.~Yoshikawa, N.~Y. Yamasaki, Y.~Suto, T.~Ohashi, K.~Mitsuda, Y.~Tawara, and
  A.~Furuzawa, ``Detectability of the warm/hot intergalactic medium through
  emission lines of \textrm{O VII} and \textrm{O VIII},'' {\em PASJ}~{\bf 55},
  p.~879, 2003.

\bibitem{yoshikawa04}
K.~{Yoshikawa}, K.~{Dolag}, Y.~{Suto}, S.~{Sasaki}, N.~Y. {Yamasaki},
  T.~{Ohashi}, K.~{Mitsuda}, Y.~{Tawara}, R.~{Fujimoto}, T.~{Furusho},
  A.~{Furuzawa}, M.~{Ishida}, Y.~{Ishisaki}, and Y.~{Takei}, ``Locating the
  warm--hot intergalactic medium in the simulated local universe,'' {\em
  PASJ}~{\bf 56}, p.~939, 2004.

\end{thebibliography}
\bibliographystyle{spiebib}   
\end{document}